\documentclass[10pt,letterpaper,twocolumn]{article} %% two column, final layout

\usepackage{ol2}
\usepackage[draft,implicit=false]{hyperref}
\usepackage{amsmath}

\begin{document}

\twocolumn[

\title{Experimental confirmation of long-memory correlations in star-wander data}

\author{Luciano Zunino,$^{1,2,*}$ Dami\'an Gulich,$^{1,3}$ Gustavo Funes,$^{1}$ and Aziz Ziad$^{4}$}

\address{
$^1$Centro de Investigaciones \'Opticas (CONICET La Plata - CIC), C.C. 3, 1897 Gonnet, Argentina\\
$^2$Departamento de Ciencias B\'asicas, Facultad de Ingenier\'ia, Universidad Nacional de La Plata (UNLP), 1900 La Plata, Argentina\\
$^3$Departamento de F\'isica, Facultad de Ciencias Exactas, Universidad Nacional de La Plata (UNLP), 1900 La Plata, Argentina\\
$^4$Laboratoire J.L. Lagrange UMR 7293, Universit\'e de Nice Sophia-Antipolis/CNRS/OCA, 06108 Parc Valrose, Nice, France
$^*$Corresponding author: lucianoz@ciop.unlp.edu.ar}

\begin{abstract}
In this letter we have analyzed the temporal correlations of the angle-of-arrival fluctuations of stellar images. Experimentally measured data were carefully examined by implementing multifractal detrended fluctuation analysis. This algorithm is able to discriminate the presence of fractal and multifractal structures in recorded time sequences. We have confirmed that turbulence-degraded stellar wavefronts are compatible with a long-memory correlated monofractal process. This experimental result is quite significant for the accurate comprehension and modeling of the atmospheric turbulence effects on the stellar images. It can also be of great utility within the adaptive optics field.
\end{abstract}
\ocis{000.5490, 010.1290, 010.1300, 010.1330, 010.7350, 350.1260.}]

\noindent
It is well-known that optical waves are strongly affected by the refractive-index fluctuations along the optical path. Because of this phenomenon, the spatial resolution of Earth telescopes is mainly limited by the atmospheric turbulence rather than by the optical design and optical quality~\cite{Roggemann1996}. This is the prime reason why the best ground-based sites, with extraordinary stable atmosphere and minimum seeing, are very carefully selected before placing any large telescope. Speckle imaging methods and adaptive optics techniques were introduced to mitigate turbulence-induced phase fluctuations. Furthermore, space telescopes have also been developed as an efficient but too expensive solution to overcome this unwanted drawback.

Performance of ground-based optical astronomy is directly linked to the atmospheric conditions. Consequently, the accurate modeling of the atmospheric turbulence effects is crucial for improving astronomical observations. For example, in adaptive optics systems, atmospherically distorted wavefront predictions could help to decrease wavefront reconstruction errors~\cite{Jorgenson1992}. Since the wavefront tilt is the dominant atmospheric aberration across the telescope pupil, its statistical characterization turns out to be of paramount importance. Atmospherically distorted images have traditionally been modeled as a fully random stochastic process~\cite{Goodman1985}. Schwartz~\textit{et al.}~\cite{Schwartz1994} enhanced this idea by identifying turbulence-degraded wavefronts as fractal surfaces. More precisely, the wavefront phase is modeled in the inertial range as a fractional Brownian motion surface with a Hurst exponent $H=5/6$. This fractal model can be also associated with the temporal behavior by assuming the validity of the frozen flow hypothesis~\cite{Schwartz1994}. Fractional Brownian motion (fBm) is a family of Gaussian self-similar stochastic processes with stationary increments (fractional Gaussian noise, fGn) widely used for modeling fractal phenomena which have empirical spectra of power-law type, $1/f^{\alpha}$ and $\alpha=2H+1$, with $1<\alpha<3$~\cite{Mandelbrot1968}. The Hurst exponent $H \in (0,1)$ quantifies their intrinsic long-range correlations: when $H>1/2$, consecutive increments tend to have the same sign so that these processes are \textit{persistent}~\cite[Chap. 9]{Feder1988}. For $H<1/2$, on the other hand, consecutive increments are more likely to have opposite signs, and it is said that the processes are \textit{anti-persistent}~\cite[Chap. 9]{Feder1988}. The standard memoryless Brownian motion (random walk) is recovered for $H=1/2$. Following a different hypothesis, Jorgenson~\textit{et al.}~\cite{Jorgenson1991} suggested that atmospherically induced effects on stellar images may be better modeled by a chaotic deterministic than by a random process. However, a few years later, the same authors concluded in favor of a correlated stochastic dynamics~\cite{McGaughey1991} in agreement with the fBm model proposed in~\cite{Schwartz1994}. It is worth mentioning that this striking memory effect has been previously confirmed in a more general framework: the propagation of optical waves through disordered media~\cite{Freund1988}.

Taking into account the ubiquity of multifractals in nature, we are looking for the presence of multiple scaling exponents in the same range of temporal scales for star-wander data. The accurate identification of these scaling exponents is fundamental to develop suitable models for simulation and forecasting purposes. In this letter the fractal and multifractal nature of experimentally recorded angle-of-arrival (AA) fluctuations of stellar images is examined via multifractal detrended fluctuation analysis (MF-DFA)~\cite{Kantelhardt2002}. This technique is particularly reliable for unveiling the fractal and multifractal scalings in experimental time series. Even though other methods have been proposed for the same purpose, MF-DFA is widely accepted due to its easy implementation and accuracy. Furthermore, it is recommended in the majority of situations in which the multifractal character of data is unknown \textit{a priori}~\cite{Oswiecimka2006}. 

MF-DFA is based on the traditional DFA method~\cite{Peng1994}, which has been widely proved to be robust, simple and versatile for accurately quantifying the long-range correlations embedded in nonstationary time series~\cite{Kantelhardt2001}. Briefly explained, given a time series $S=\{x_t,t=1,\dots,N\}$, with $N$ being the number of observations, the cumulated data series $Y\left(i\right)=\sum_{t=1}^{i}\left(x_{t}-\langle x \rangle \right)$, with $i=1,\dots,N$ and $\langle x \rangle = \left(\sum_{t=1}^{N} x_t\right)/N$, is considered. This profile is divided into $\left\lfloor N/s \right\rfloor$ nonoverlapping windows of equal length $s$ ($\left\lfloor a \right\rfloor$ denotes the largest integer less than or equal to $a$). A local polynomial fit $y_{\nu,\,m}\left(i\right)$ of degree $m$ is fitted to the profile for each window $\nu=1,\dots,\left\lfloor N/s \right\rfloor$. The degree of the polynomial can be varied to eliminate constant ($m=0$), linear ($m=1$), quadratic ($m=2$) or higher order trends of the profile. Then the variance of the detrended time series is evaluated by averaging over all data point $i$ in each segment $\nu$, $F^{2}_m\left(\nu,s\right)=\frac{1}{s}\sum_{i=1}^{s}\left\{Y\left[\left(\nu-1\right)s+i\right]-y_{\nu,\,m}\left(i\right)\right\}^{2}$, for $\nu=1,\dots,\left\lfloor N/s \right\rfloor$. In order to analyze the influence of fluctuations of different magnitudes and on different time scales, the generalized $q\text{th}$ order fluctuation function given by
$F_{q}\left(s\right)=\left\{\frac{1}{\left\lfloor N/s \right\rfloor}\sum_{\nu=1}^{\left\lfloor N/s \right\rfloor}\left[F^{2}_m\left(\nu,s\right)\right]^{q/2}\right\}^{1/q}$ is estimated for different values of the time scale $s$ and for different values of the order $q$ ($q \neq 0$). When $q=0$ a logarithmic averaging procedure has to be employed because of the diverging exponent. For $q=2$, the conventional fractal DFA algorithm is retrieved. Generally, if the time series $S=\{x_t,t=1,\dots,N\}$ has long-range power-law correlations, $F_{q}\left(s\right)$ scales with $s$ as
\begin{equation}
F_{q}\left(s\right)\sim s^{h\left(q\right)}
\label{powerlaw}
\end{equation}
for a certain range of $s$. The scaling exponents $h(q)$, usually known as generalized Hurst exponents, are estimated by analyzing the double logarithmic plot of $F_{q}\left(s\right)$ versus $s$ for each value of $q$. Ideally, if the series is monofractal and stationary, then $h(q)$ is constantly equal to the Hurst exponent $H$, i.e. independent of $q$ ($h(q)=H$). Otherwise, a multifractal structure is observed when the scaling behaviors of small and large fluctuations are different. In this case the generalized Hurst exponent is a decreasing function of $q$ and the main Hurst exponent can be estimated from the second moment ($h(2)=H$). The generalized Hurst exponents with negative order $q$ describe the scaling of small fluctuations because the segments $\nu$ with small variance will dominate the average $F_{q}\left(s\right)$ for this $q$-range. On the contrary, for positive order $q$ the windows $\nu$ with large variance have stronger influence and, thus, the scaling of large fluctuations is examined. The strength of multifractality present in data is usually defined as the spread of the generalized Hurst exponents~\cite{Grech2013}. As small fluctuations are characterized by larger scaling exponent than those associated with large fluctuations, $h(q)$ for $q<0$ are larger than those for $q>0$, and the multifractality degree can be quantified by
\begin{equation}
\Delta h \equiv h(-q)-h(q)
\label{multifractality}
\end{equation}
for a larger value of the moment $q$. For further details about MF-DFA and its implementation in Matlab we recommend~\cite{Ihlen2012}.

\begin{figure}[h]
\centerline{\includegraphics[width=7.5cm]{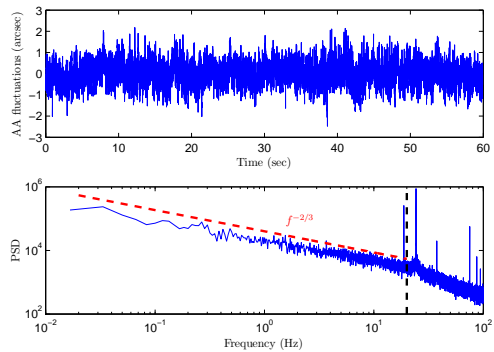}}
\caption{Representative sample of the AA fluctuations (top) and average PSD of the nineteen sequences of real wavefront slopes (bottom). The theoretical expected $-2/3$ power law behavior at the low-frequency regime is plotted (red dashed line). The knee frequency is also indicated (vertical black dashed line). Peaks observed at high frequencies are due to vibrations in the experimental arrangement.}
\label{figure1}
\end{figure}

The experimental AA fluctuation measurements were taken by the Generalized Seeing Monitor (GSM) instrument~\cite{Ziad2000,Ziad2012} on a star at Paranal Observatory (Antofagasta, Chile). More precisely, nineteen independent sets of data recorded on December 16, 2007 were carefully analyzed. The AA fluctuations are measured with a tight and regular sampling of 5 ms during approximately 1 min acquisition time (time series length $N = 11,984$). The data acquisition is repeated typically every 4 min. Figure~\ref{figure1} shows one representative sample of the AA fluctuations (top plot) together with the average temporal power spectral density (PSD) of the nineteen sequences (bottom plot). The expected $-2/3$ power law scaling at the low-frequency region is shown (red dashed line). Vertical black dashed line indicates the knee frequency that appears due to the spatial averaging over the telescope aperture~\cite{Conan1995}.

\begin{figure}[h]
\centerline{\includegraphics[width=7.5cm]{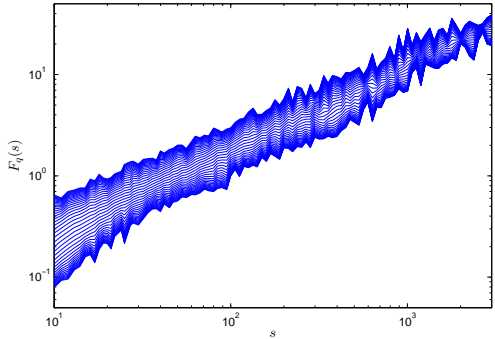}}
\caption{Fluctuation functions $F_q(s)$ as a function of the scale $s$ for the AA fluctuations plotted in Fig.~\ref{figure1}. A detrending polynomial of order $m=2$ and 100 different scales $s \in [10,N/4]$ equally spaced in the logarithmic scale were employed in the MF-DFA implementation. The order $q$ ($q=-20,-19,...,20$) increases from bottom to top. The behavior observed is representative for the whole data set.}
\label{figure2}
\end{figure}

We have analyzed the fractal and multifractal behavior of the AA fluctuations of stellar images by implementing the MF-DFA technique with a detrending polynomial of second order $m=2$. Similar results were obtained with other orders of the detrending polynomials ($m=1$, $3$ and $4$). One hundred time scales $s \in [10,N/4]$ equally distributed in the logarithmic scale were selected for estimating the fluctuation functions. We restrict the moment $q$ to the range $[-20,20]$ with step equal to 0.25 ($q=-20,-19.75,...,20$). As an illustrative example, fluctuation functions $F_q(s)$ for the AA fluctuations plotted in Fig.~\ref{figure1} are shown in Fig.~\ref{figure2}. Only $F_q(s)$ with integer moment, i.e. $q=-20,-19,...,20$, are depicted for a better visualization. From this figure it can be concluded that the slope of the fluctuation functions in the log-log plot, $h(q)$, slightly decrease with the moment $q$. 

\begin{figure}[b]
\centerline{\includegraphics[width=7.5cm]{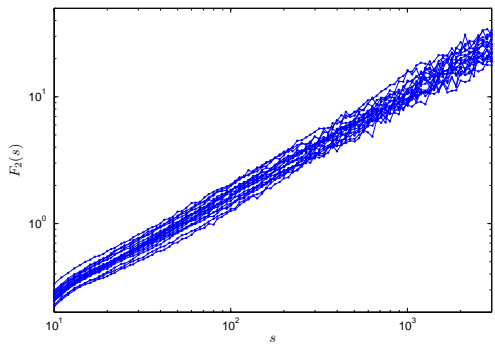}}
\caption{Fluctuation functions $F_2(s)$ as a function of the scale $s$ for the nineteen independent sets of AA fluctuations. The slope of the best linear fit obtained for each one of these fluctuation functions is the Hurst exponent estimator (standard DFA technique~\cite{Peng1994}) of the experimental records.}
\label{figure3}
\end{figure}

To better understand the fractal nature, Fig.~\ref{figure3} shows the fluctuation function for the second moment, $F_2(s)$, as a function of the scale $s$ for the nineteen independent sets of AA fluctuations. It should be emphasized the excellent linearity observed for all the time scales. This finding allows to confirm the existence of a well-defined power law behavior, $F_{2}\left(s\right)\sim s^{h\left(2\right)}=s^{H}$, and, accordingly, a fractal dynamics, in the full analyzed range. 

\begin{figure}[h]
\centerline{\includegraphics[width=7.5cm]{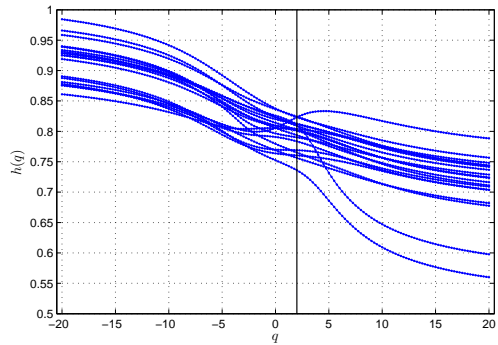}}
\caption{Generalized Hurst exponents $h(q)$, estimated in the full fitting range $s \in [10,N/4]$, as a function of the order $q$ for the nineteen independent sets of AA fluctuations. Vertical black continuous line indicates the estimated values for the main Hurst exponent ($H=h(2)$).}
\label{figure4}
\end{figure}

\begin{figure}[b]
\centerline{\includegraphics[width=7.7cm]{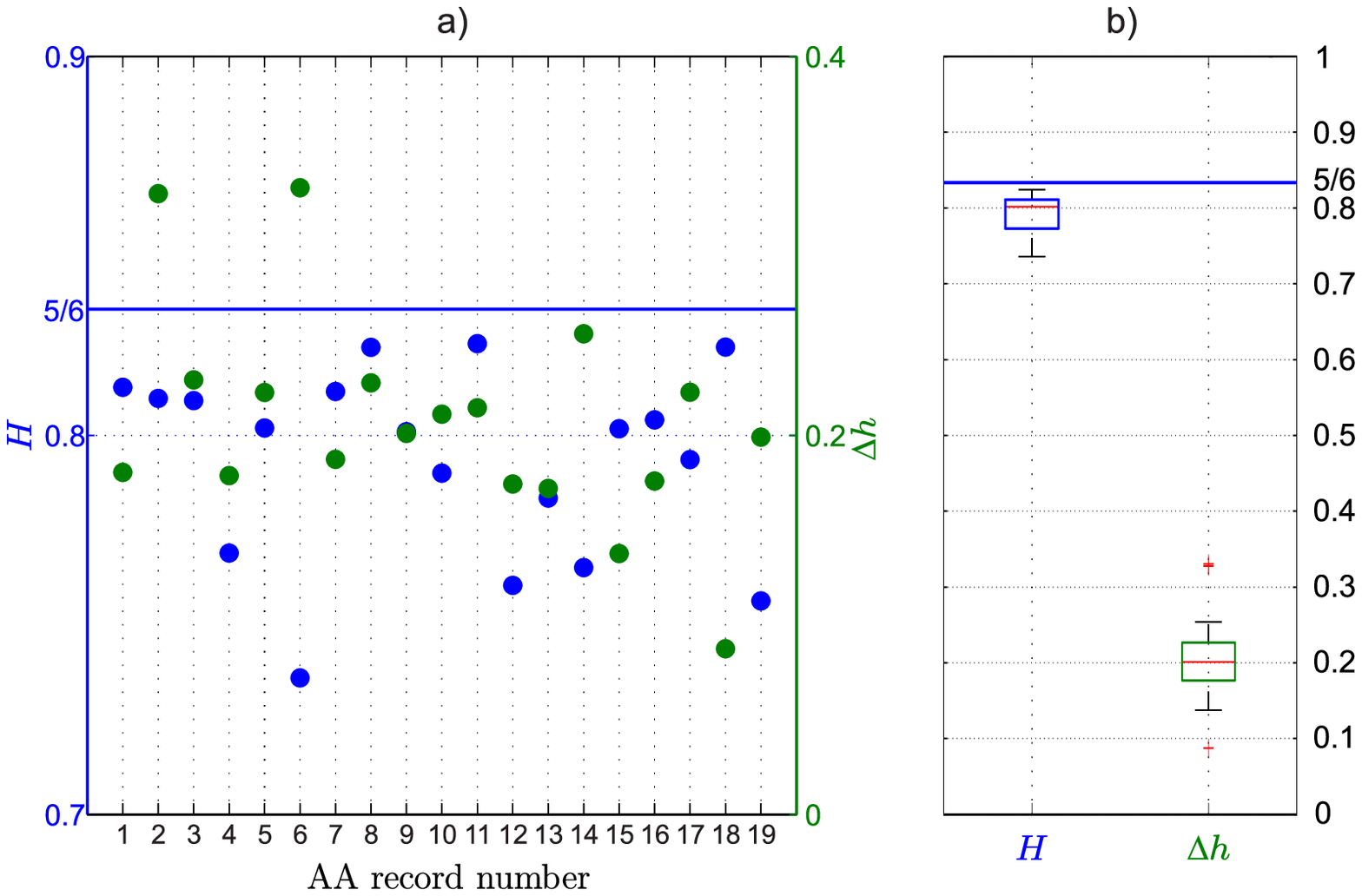}}
\caption{a) Estimated values for the Hurst exponent $H$ (blue dots) and multifractality degree $\Delta h$ (green dots) for the nineteen independent experimental sets of AA fluctuations. b) Related boxplots for both quantifiers. The theoretical expected value for the Hurst exponent within the Kolmogorov model ($H=5/6$) is indicated (horizontal blue continuous lines).}
\label{figure5}
\end{figure}

Generalized Hurst exponents estimated in the full time scale range, i.e. fitting range $s \in [10,N/4]$, for the nineteen independent sets of AA fluctuations are plotted in Fig.~\ref{figure4}. Main Hurst related exponents ($H=h(2)$) are indicated with a vertical black continuous line. Values estimated for this parameter ($H$) together with those obtained for the multifractal strength ($\Delta h$), defined according to Eq.~(\ref{multifractality}), are detailed in Fig.~\ref{figure5}. Mean and standard deviation of the estimators of both quantities, namely Hurst exponent $H$ and multifractality strength $\Delta h$, over the whole data set are $0.79 \pm 0.03$ and $0.21 \pm 0.06$, respectively. On the one hand, a persistent stochastic behavior is concluded from the DFA analysis. On the other hand, the results obtained for the generalized Hurst exponents suggest a small degree of multifractality. This small spread of the $h(q)$ values can be directly ascribed to finite-size effects. More precisely, an apparent, false, multifractality degree $\Delta h \approx 0.2$ is commonly found in purely long-range correlated monofractal signals~\cite{Schumann2011}. As it has been proved by Grech and Pamu{\l}a~\cite{Grech2013}, this spurious effect appears as a result of finite length of analyzed data and is additionally amplified by the presence of long-term memory. In order to better clarify this issue, we have estimated the generalized Hurst exponents of one hundred independent realizations of fGn with Hurst exponent $H=0.8$. These numerical simulations, with the same length $N$ of the AA fluctuation time series, were generated via the function \textit{wfbm} of MATLAB. This algorithm simulates fBm following the method proposed by Abry and Sellan~\cite{Abry1996}. The fGn numerical realizations are obtained through successive differences of the fBm simulations. MF-DFA with the same parameters used for the AA fluctuation records was implemented for this numerical study. Mean and standard deviation of the estimated values for $H$ and $\Delta h$ are $0.80 \pm 0.02$ and $0.16 \pm 0.03$, respectively. These results confirm the existence of a spurious multifractality in monofractal long-range correlated time series due to finite-size effects. Consequently, AA fluctuations of stellar images can be modeled, at least in a first approximation, as a monofractal long-memory correlated stochastic process.

Our experimental results support the fBm model for the atmospherically induced wavefront degradations proposed by Schwartz~\textit{et al.}~\cite{Schwartz1994}. The estimated Hurst exponent, however, is always below the $5/6$ value expected for a conventional Kolmogorov theory. This smaller Hurst exponent can be ascribed to a non-Kolmogorov behavior of the atmospheric turbulence because there exist evidence of deviations from the Kolmogorov model in the upper atmosphere~\cite{Dayton1992,Nicholls1995}. Indeed, Du~\textit{et al.}~\cite{Du2010} have theoretically found that the power law of the temporal power spectra of AA fluctuations for low frequencies is modified when a generalized power spectrum model for the refractive-index fluctuations, i.e. non-Kolmogorv turbulence, is considered. This change in the scaling law for the low-frequency regime can be directly associated with the deviations from the Kolmogorov-expected Hurst exponent that we have experimentally observed.

Summarizing, we have confirmed the presence of long-range correlations in AA fluctuations of stellar wavefronts propagating through atmospheric turbulence. The estimated Hurst exponent is always near but below the theoretically expected $5/6$ value for a Kolmogorov turbulence. Indeed, this smaller estimated Hurst exponent can be understood in terms of a non-Kolmogorov turbulence model. It is worth emphasized that these results allow to suggest that the turbulence-degraded wavefront phase can be modeled as a fBm with $H \approx 0.8$. The inherent predictability associated with this persistent stochastic process might be useful to improve the performance of high-angular-resolution techniques. Further analysis with a larger database are planned for optimizing the Hurst exponent estimation.

This work was partially supported by Consejo Nacional de Investigaciones Cient\'ificas y T\'ecnicas (CONICET), Argentina and Universidad Nacional de La Plata (UNLP), Argentina.

\pagebreak

\end{document}